\documentclass{optica-article}
\usepackage{soul}

\journal{opticajournal} 

\articletype{Research Article}

\usepackage{lineno}
\usepackage{siunitx}

\begin{document}

\title{Optimizing probes for multi-beam ptychography}

\author{Runqing Yang,\authormark{1,*} Pablo Villanueva-Perez,\authormark{1} Maik Kahnt\authormark{2}}

\address{\authormark{1}Division of Synchrotron Radiation Research and NanoLund, Department of Physics, Lund University, Lund, 22100, Sweden\\
\authormark{2}MAX IV Laboratory, Lund University, Box 118, 221 00, Lund, Sweden}

\email{\authormark{*}runqing.yang@maxiv.lu.se} 


\begin{abstract*} 

Multi-beam ptychography (MBP) offers a scalable solution to improve the throughput of state-of-the-art ptychography by increasing the number of coherent beams that illuminate the sample simultaneously.
However, increasing the number of beams in ptychography makes ptychographical reconstructions more challenging and less robust.
It has been demonstrated that MBP reconstructions can be made more robust by using well-structured and mutually separable probes. 
Here, we present a quantitative framework to assess probe sets based on separability, uniformity, and fabrication feasibility. 
We show that Hadamard-based binary phase masks consistently outperform Zernike polynomials, experimentally feasible phase plates, and spiral phase masks across varying scan densities. 
While spiral masks yield comparable resolution, they scale less efficiently due to increased structural complexity. 
Our results establish practical criteria for evaluating and designing structured probes to enable more robust and scalable implementation of MBP in high-throughput coherent X-ray and EUV imaging.

\end{abstract*}

\section{Introduction}

Ptychography~\cite{rodenburg2008ptychography, pfeiffer2018x} is a widely used technique in coherent X-ray imaging~\cite{miao2011coherent}.
It enables high-resolution, quantitative phase-contrast imaging by scanning a sample with a small probing beam with overlap between the adjacent scan points and recording the resulting diffraction patterns in the far field.  
Using these diffraction patterns, it iteratively reconstructs both the complex-valued object transmission and the probe function itself with high resolution.
Compared to conventional imaging methods, ptychography is particularly well-suited for weakly scattering or transparent specimens, owing to its phase sensitivity and quantitativeness~\cite{dierolf2010ptychographic}. 
This makes it a valuable tool for applications ranging from materials science~\cite{michelson2022three, aidukas2024high, das2024insitucatalysts} to biology~\cite{marrison2013ptychography, bosch20233d}, where subtle phase variations can encode critical structural information. 
These applications often demand high sensitivity, large fields of view, or high throughput~\cite{pan2020high, Reinhardt2017beamstoplowscattering, takahashi2013high, okawa2024sensitivity}, motivating the development of more scalable ptychographic methods. 

However, as a scanning-based technique, conventional ptychography, also known as single-beam ptychography (SBP), is limited by the need to scan, long acquisition times, and the requirement of a highly coherent and high-intensity probing beam.
These limitations are especially pronounced for large-area, high-resolution, or high photon energy X-ray applications where the coherent fraction of current X-ray sources is low and efficient photon use is critical. 
To address these challenges, multi-beam ptychography (MBP) ~\cite{batey2014information, bevis2016multiple, bevis2018multiple} has been developed to illuminate multiple regions of the sample simultaneously by splitting the partially coherent primary beam into multiple coherent in themselves, but mutually incoherent beams that illuminate the sample in multiple spots simultaneously.
Utilizing previously unused photons allows for increasing both throughput and photon utilization~\cite{yao2020multi}. 
Recent experimental and computational studies have demonstrated the feasibility and promise of MBP ~\cite{lyubomirskiy2022multi, aastrand2024adaptive, bevis2018multiple, li2024x,  wittwer2021upscaling, hirose2020multibeam}, yet key questions related to optimized probe design and information separation remain open.

In SBP, it is well established that the choice of the probe strongly influences the quality and robustness of the phase retrieval. 
Various studies have shown that well-tailored structured illuminations can significantly improve reconstruction stability and resolution~\cite{odstrvcil2019towards, eschen2024structured, guizar2012role, ji2022resolution}.
However, MBP introduces an additional layer of complexity.
Not only must each individual beam itself have optimal properties for SBP, but the multiple beams must also be sufficiently distinct from one another to ensure that the contributions from the respective beams to the recorded diffraction patterns can be uniquely disentangled during reconstruction~\cite{lyubomirskiy2022multi, li2024x, penagos2025multiplexing, wengrowicz2024unsupervised}. 
Recent implementations have explored different approaches to generate multiple structured beams, including the use of customized phase plates ~\cite{lyubomirskiy2022multi,li2024x}, Fresnel zone plate with partial zone inversions ~\cite{aastrand2024adaptive, aastrand2023multi}, and arrays of structured apertures ~\cite{penagos2025multiplexing, eschen2024structured}. 
While these methods demonstrate the feasibility of creating diverse multi-beam configurations, how to systematically design and evaluate beams that remain individually well-suited for phase retrieval while being sufficiently distinct for reliable multi-beam reconstruction had not been systematically studied yet~\cite{lyubomirskiy2022multi}.

To enable robust and practical implementation of MBP, it is essential to design probe sets that jointly satisfy three critical criteria: separability, uniformity, and feasibility. 
Separability ensures that individual probes can be distinguished and reconstructed independently with minimal cross-talk, enabling accurate phase retrieval even when multiple beams illuminate the sample simultaneously.
Uniformity requires that all beams exhibit comparable size, amplitude distributions, structural features, and reconstruction fidelity, so that no single beam dominates or limits the overall image quality or required scanning parameters.
While these two factors primarily impact reconstruction performance, feasibility plays an equally crucial role in translating conceptual designs into practical experimental tools. 
We define feasibility in this context as encompassing three aspects:
(i) scalability, meaning the design should allow the generation of any number of beams without introducing prohibitive optical complexity;
(ii) photon efficiency, ensuring that the available flux is used effectively across all beams to maintain signal levels;
And (iii) manufacturability, meaning the probe structures should be compatible with existing optical fabrication technologies. 
Meeting all three feasibility requirements is particularly important for high-throughput or low-dose imaging applications, where both performance and implementation constraints must be balanced.

In this work, we investigate how different probe designs affect these three factors and how design choices can be optimized to jointly maximize separability, uniformity, and feasibility.

\section{Probe design}

To translate the general principles of MBP into practical experimental implementations, the probe design needs to meet the requirements of separability, uniformity, and feasibility, as outlined in the previous section. 

Our design strategy aims to fulfill these constraints by combining structured phase modulation with experimentally realistic configurations.
We employ square apertures in the beam optics plane, which can be arranged in a dense tileable grid \cite{lyubomirskiy2022multi,li2024x}. 
This configuration partitions a larger illumination on the beam forming optics into multiple distinct and in themselves coherent beams while maintaining high photon throughput. 
To structure each probe independently, we modulate the pupil function (the aperture function of the beam forming elements), a method compatible with existing X-ray and EUV optical systems.
We choose phase-only modulation to ensure photon efficiency: in the X-ray regime, phase shifts can be readily achieved with thin phase plates or diffractive optics~\cite{levitan2025single, marchesini2019shaping}, while for EUV wavelengths, absorption masks are more straightforward to fabricate, but phase modulation remains feasible and helps maximize photon use \cite{eschen2024structured}. 

Building on these design principles, we investigated four groups of pupil functions to generate structured and mutually distinct probes. 
First, we considered two-dimensional Zernike polynomials adapted for square apertures\cite{ye2014comparative, odstrvcil2019towards}. 
Second, we explored Hadamard matrix-like binary phase modulations\cite{hadamard1893resolution}, which are practical to fabricate and inspired by orthogonal matrix designs at the pupil level, expected to promote well-separable beam modes in the object plane (see Algorithm S1 in the Supplementary Document for generation details).
In addition, we included phase plate designs that have already demonstrated experimental feasibility in previous X-ray implementations\cite{lyubomirskiy2022multi, li2024x}, providing a reliable baseline for practical fabrication. 
Finally, we examined spiral phase masks similar to absorbing masks applied in EUV MBP\cite{penagos2025multiplexing}, which vary the number of spiral arms and rotation direction to structure multiple beams. 

\begin{figure}[h]
    \centering
    \includegraphics[width=\textwidth]{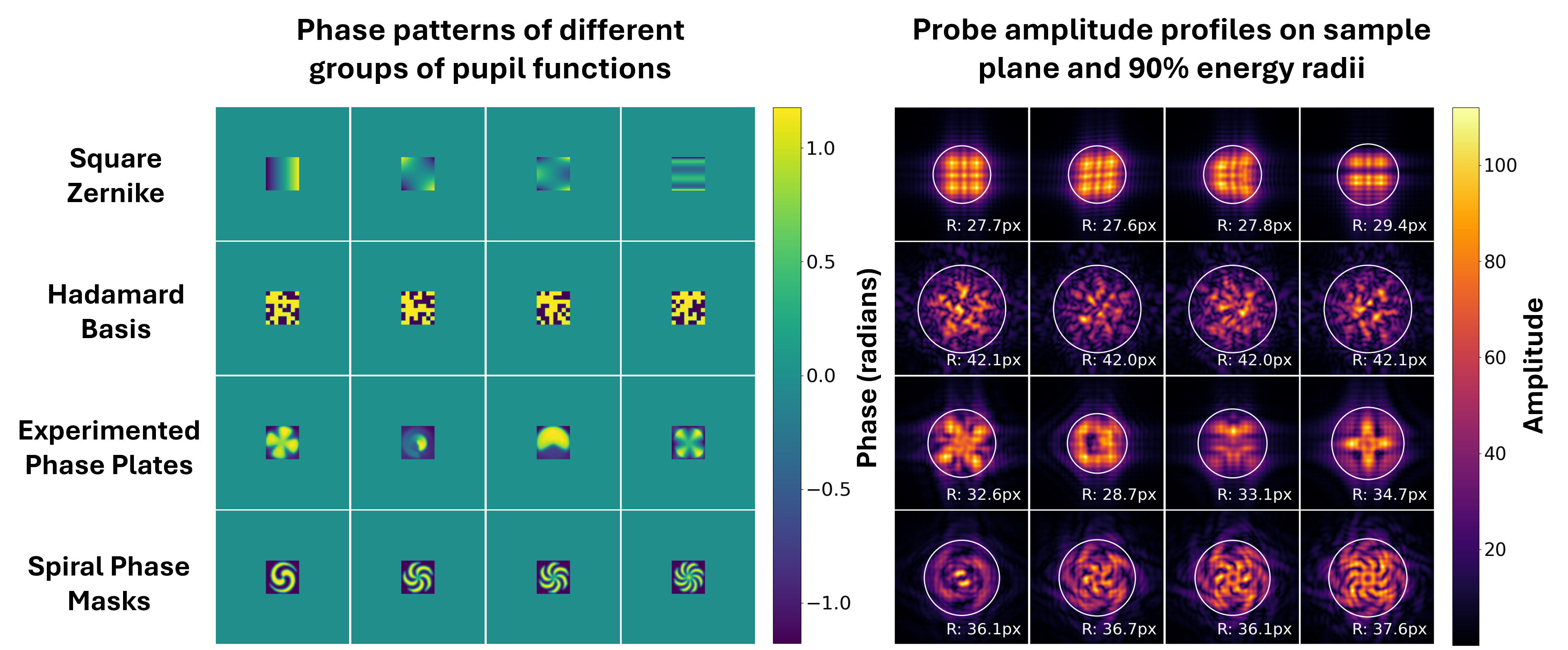} 
    \caption{Representative examples of four distinct wavefront modulation designs used for probe generation. The left panels show the phase distributions of the corresponding pupil functions, while the right panels display the resulting probe amplitudes after propagating to the sample. The effective probe radius, defined as the radius enclosing 90\% of the total intensity, is quantified by cumulative intensity analysis, indicated by the white circles and annotated with the corresponding radius values in pixels.}
    \label{fig:probes}
\end{figure}

This diverse set of probe design strategies enables a systematic comparison of how different phase modulation schemes perform under the combined demands of MBP: high separability, uniformity, and feasibility.
All resulting pupil functions were numerically propagated to simulate the corresponding probe amplitude distributions at the sample plane. 
Fig.~\ref{fig:probes} illustrates representative examples of the designed pupil phases and their resulting propagated probe amplitudes at the sample plane, including an indication of the effective radius containing \SI{90}{\percent} of the intensity. 

The simulated probes were generated using parameters representative of feasible experimental conditions. 
Each complex-valued pupil function was defined on a 128 x 128 pixel grid with a constant amplitude inside a central square pupil aperture of 32 x 32 pixels and an amplitude of zero outside that central square. 
The central apertures of the pupil functions were only structured through the different distinctive phase patterns.
The maximum relative phase shifts induced was chosen as \(+\frac{3\pi}{8}\) and \(-\frac{3\pi}{8}\) radians across all structuring options. 
This phase selection results in a total phase variation of \(\frac{3\pi}{4}\), which provides sufficient phase diversity.
To obtain the probe distributions at the sample plane, the pupil functions were first Fourier transformed to obtain the wave field at the focal plane and then propagated via the Fresnel formalism~\cite{maiden2009improved} over a \SI{3}{\milli\meter} distance to the sample plane to simulate typical experimental defocus conditions.

\section{Object design}
To enable a fair assessment of single-beam and multi-beam ptychographic performance, we constructed a synthetic test object that imposes an unbiased and consistent reconstruction challenge across all probe configurations and any randomly selected sub-region of the object. 
The design incorporates distinct regions spanning a range of structural complexity, including featureless areas, pure phase modulations, pure amplitude variations, and regions that combine both effects. 
This diversity ensures that no single probe is inadvertently favored or disadvantaged during reconstruction due to the sub-region of the sample it gets to image.

To meet the requirements for statistical uniformity and spectral consistency, we generated two independent arrays based on Perlin noise fields of different feature sizes~\cite{Abdolhoseini2019Neuron}. 
The complex transmission function of the synthetic object was defined as 
\( T(x,y) = A(x,y) \cdot e^{i \phi(x,y)} \), where both amplitude 
\( A(x,y) \) and phase \( \phi(x,y) \) were independently generated using 
Perlin noise fields with matched spatial frequency content but different seeds. 
The amplitude \( A(x, y) \) was treated as a normalized transmission factor, where \( 0 \leq A(x, y) \leq 1 \), without explicitly modeling absorption or optical density.
This approach ensures matched spectral content while maintaining statistical independence between amplitude and phase. 
By superimposing multiple Perlin noise layers with varying spatial frequencies, we ensured the presence of structures ranging from coarse to fine scales. 
A chunk-based generation strategy results in seamless tiling with non-repeating elements and thus infinite scaling.
This allows the creation of synthetic test objects of any size with spectral homogeneity across the entire field of view.

In addition to spectral and spatial considerations, we constrained the complex transmission values of the synthetic object.
Specifically, the amplitude was bounded between 0.7 and 1.0 to reflect moderate absorption, as commonly observed in weakly absorbing samples under X-ray illumination.
The phase distribution was scaled to a maximum shift of \(\frac{\pi}{2}\), providing sufficient phase contrast for evaluating reconstruction performance without introducing phase-wrapping effects.

\begin{figure}[htbp]
    \centering
    \includegraphics[width=\textwidth]{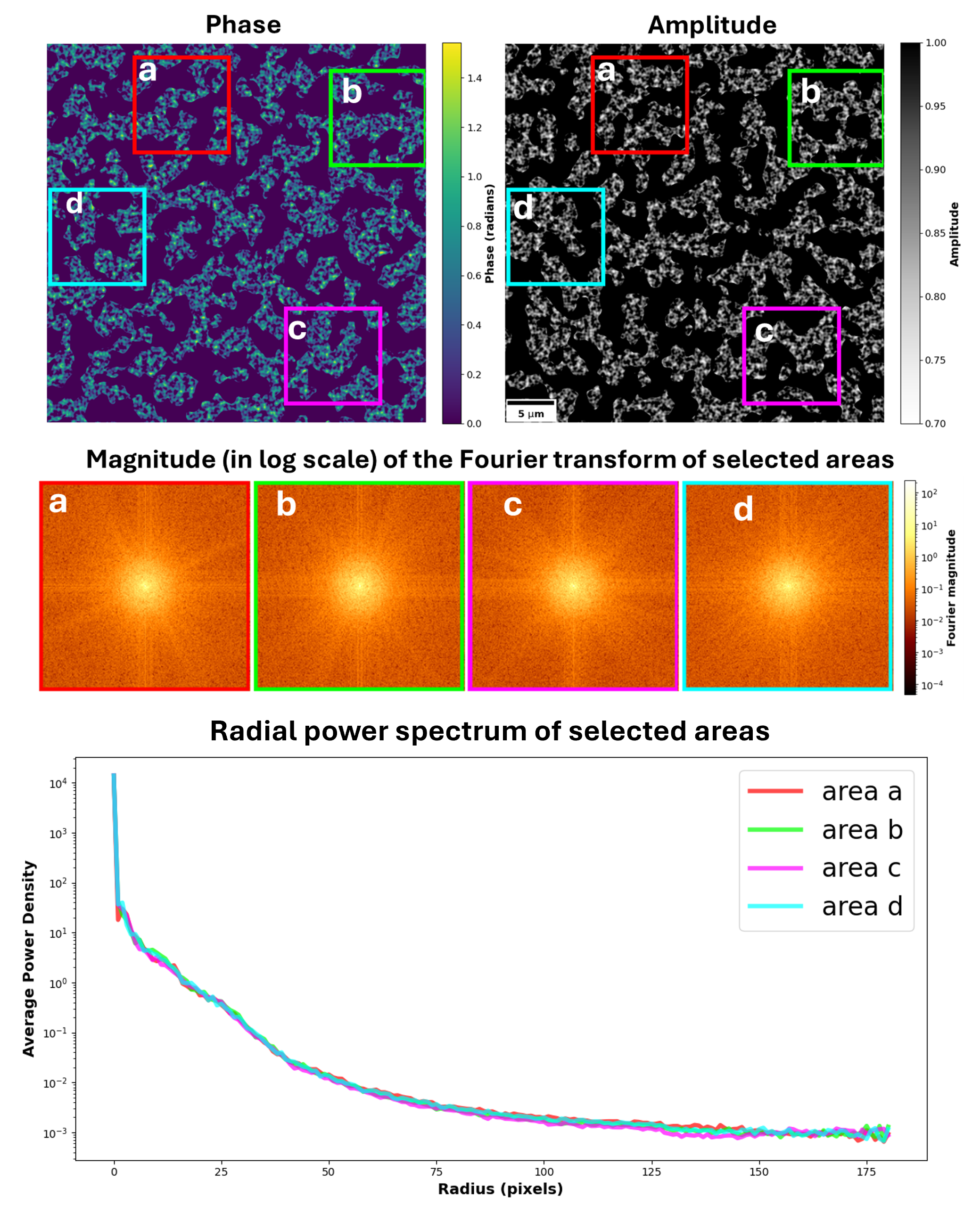} 
    \caption{
    Illustration of the synthetic test object used for probe performance evaluation. 
    Phase map and amplitude map generated using independent Perlin noise fields with identical spectral content, but different seeds for the random number generator. 
    Four randomly selected subregions (labeled a-d) are marked on the maps. 
    Below are the spectral magnitudes (in logarithmic scale) of these subregions and their corresponding radial power spectra.}
    \label{fig:object}
\end{figure}

This object design provides a robust benchmark for systematic probe evaluation under controlled yet realistic imaging conditions~\cite{Zhang2019Fourier}. 
Figure~\ref{fig:object} illustrates the constructed phase and amplitude components of the test object, along with the logarithmic Fourier magnitudes of four randomly selected representative sub-regions to demonstrate its spectral uniformity and spatial characteristics.

To validate the statistical uniformity of the synthetic object in practice, we evaluated whether probe performance comparisons are influenced by local variations in object complexity or contrast. 
A uniform object reconstructability is essential for ensuring consistent and fair evaluation of probe designs. 
In this assessment, a single representative probe from the Hadamard basis set was used to scan 16 spatially separated sub-regions of the larger test object, each with a fixed size of 512 by 512 pixels. 
Each sub-region was scanned at multiple step sizes to generate datasets with varying probe-overlap ratios between adjacent positions, and consequently, the number of diffraction patterns for the same field of view is different. 
Reconstructions were performed independently for each sub-region and each step size, and the reconstruction quality was quantified using the Fourier Ring Correlation (FRC)~\cite{van2005fourier} against the known ground truth. 
For each reconstruction, the FRC was computed between the retrieved complex-valued object and the ground truth, and the 1-bit threshold criterion~\cite{van2005fourier} was applied to determine the cutoff frequency as a statistically rigorous indicator of resolution. 
For each scan position density, the FRC cutoff frequencies from all 16 areas were averaged, and their standard deviation was computed to assess the consistency across the sub-regions of the larger test object.

\begin{figure}[h]
    \centering
    \includegraphics[width=\linewidth]{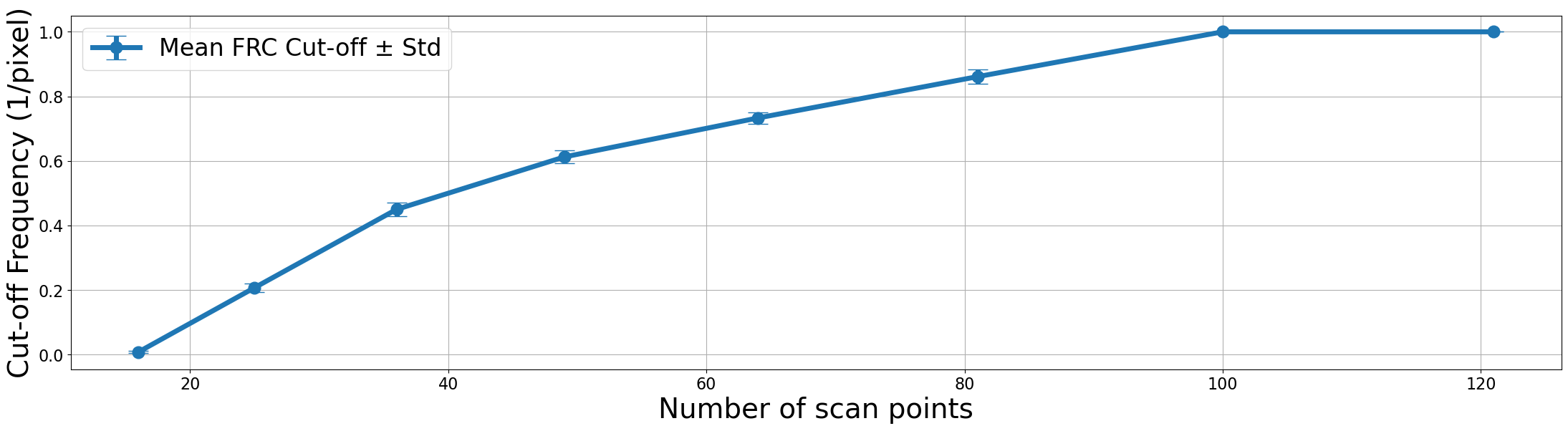}
    \caption{Obtained average reconstruction quality and standard deviation as a function of number of scan points per area, for 16 different sub-regions of the same object using a single representative probe from the Hadamard-like set.
    }
    \label{fig:dif_area}
\end{figure}

As shown in Fig.~\ref{fig:dif_area}, reconstruction resolution steadily improves with increasing scan point density (smaller step size), as expected.
More importantly, the small standard deviation across reconstructions from different sub-regions indicates minimal variability, confirming that the object structure is statistically uniform at the scale relevant to probe evaluation.
Together, these results confirm that any sub-region of the larger test object works equally well for SBP. 
Used for MBP, any observed changes in reconstruction quality can thus be attributed to the chosen set of probes and not the choice of sub-region for each probe.

\section{Probe performance}

To systematically evaluate the effectiveness of different probe design strategies, we conducted a series of numerical ptychographic simulations under consistent experimental conditions. 
All simulations used a fixed wavelength of \SI{0.953}{\angstrom}, a sample-to-detector distance of \SI{4}{\meter}, and a detector pixel size of \SI{75}{\micro\meter}. 
Each probe was assigned a total photon budget of \SI{1e7}{photons} per scan point~\cite{maiden2009improved}, and Poisson noise~\cite{thibault2012maximum} was applied to the far-field diffraction intensities to emulate realistic photon-counting conditions. 
Reconstructions were performed using the open-source package \texttt{Ptypy}~\cite{enders2016computational}, using the simulated probe as the initial estimate to avoid phase ambiguities. 
These conditions ensure that all comparisons are performed under consistent conditions and reflect intrinsic differences in probe behavior.

\subsection{Evaluating Probe Uniformity}

A key requirement for high-fidelity MBP is that all constituent beams exhibit consistent spatial characteristics. 

All pupil functions used in this study share the same size but differ in their phase structure, making them potentially mutually separable.
As a result, the corresponding object-plane probes exhibit notable variations in amplitude distribution, spatial extent, and effective beam size after propagation.
Such variations may compromise illumination uniformity and ultimately affect reconstruction quality in MBP, as overlap ratios between adjacent positions might differ for different probes. 
To quantitatively evaluate these characteristics, we simulated 12 probes for each of the four candidate design strategies introduced earlier:
(1) two-dimensional Zernike polynomials adapted to square apertures,
(2) binary phase modulations based on Hadamard matrices,
(3) phase plate designs with demonstrated experimental feasibility, 
and (4) spiral phase functions. 
The resulting amplitude profiles at the object plane were analyzed to assess spatial consistency and overall beam quality.
Beam size was estimated by measuring the radius of the smallest circle that contains 90\% of the overall intensity.

\begin{figure}[htbp]
    \centering
    \includegraphics[width=\textwidth]{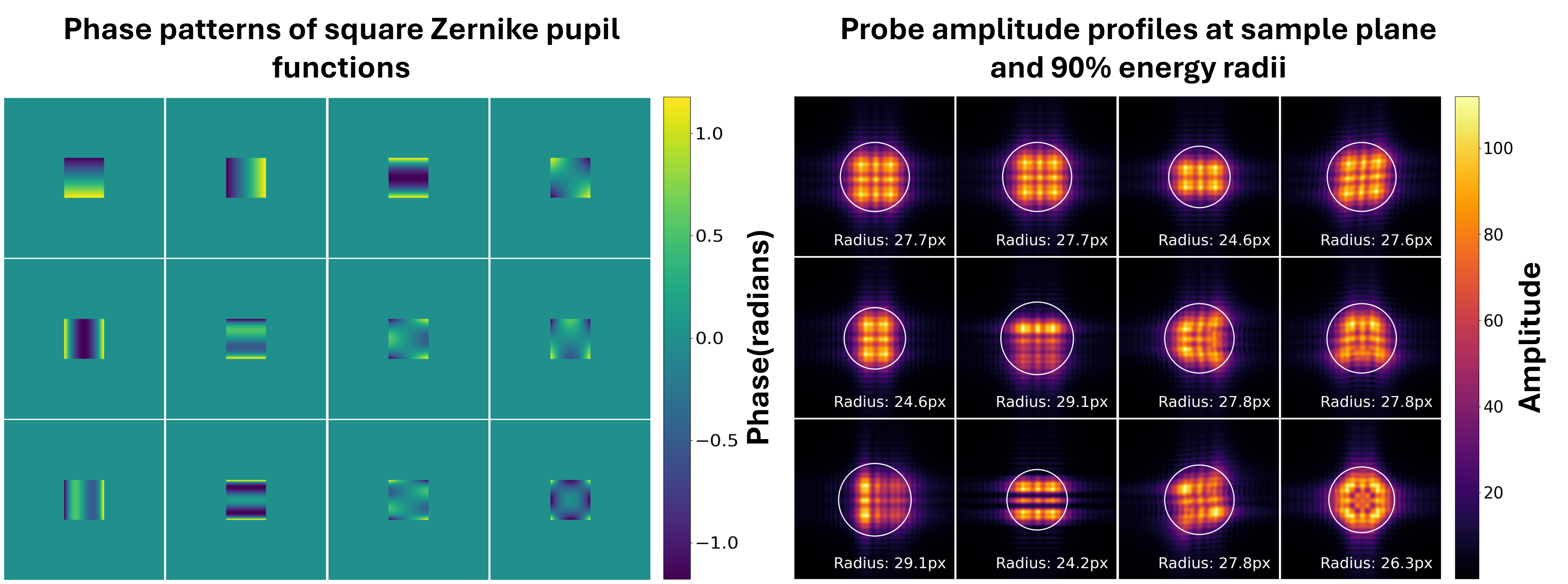}
    \caption{
    Representative examples of probes generated using square-aperture Zernike polynomials.
    Left: pupil function phase maps. Right: resulting probe amplitudes after propagation to the object plane. 
    Each probe is overlaid with a white circle indicating the effective radius enclosing 90\% of the total intensity. 
    }
    \label{fig:zernike}
\end{figure}

The amplitude distributions of the tested Zernike-based probes exhibit moderate variation in effective radius, ranging from 24.2 to 29.1 pixels, with an average of 27.1 pixels and a standard deviation of 1.61 pixels. 
While many probes maintain a reasonable degree of radial symmetry, several show asymmetric or localized amplitude distributions. 
On the other hand, some pairs exhibit nearly indistinguishable amplitude patterns, potentially making it more difficult to disentangle their contributions in the MBP reconstruction process. 
Although the Zernike approach is scalable and results in differently structured probes, its spatial diversity and consistency are limited by this geometric mismatch. (see Fig.~\ref{fig:zernike}).

\begin{figure}[htbp]
    \centering
    \includegraphics[width=\textwidth]{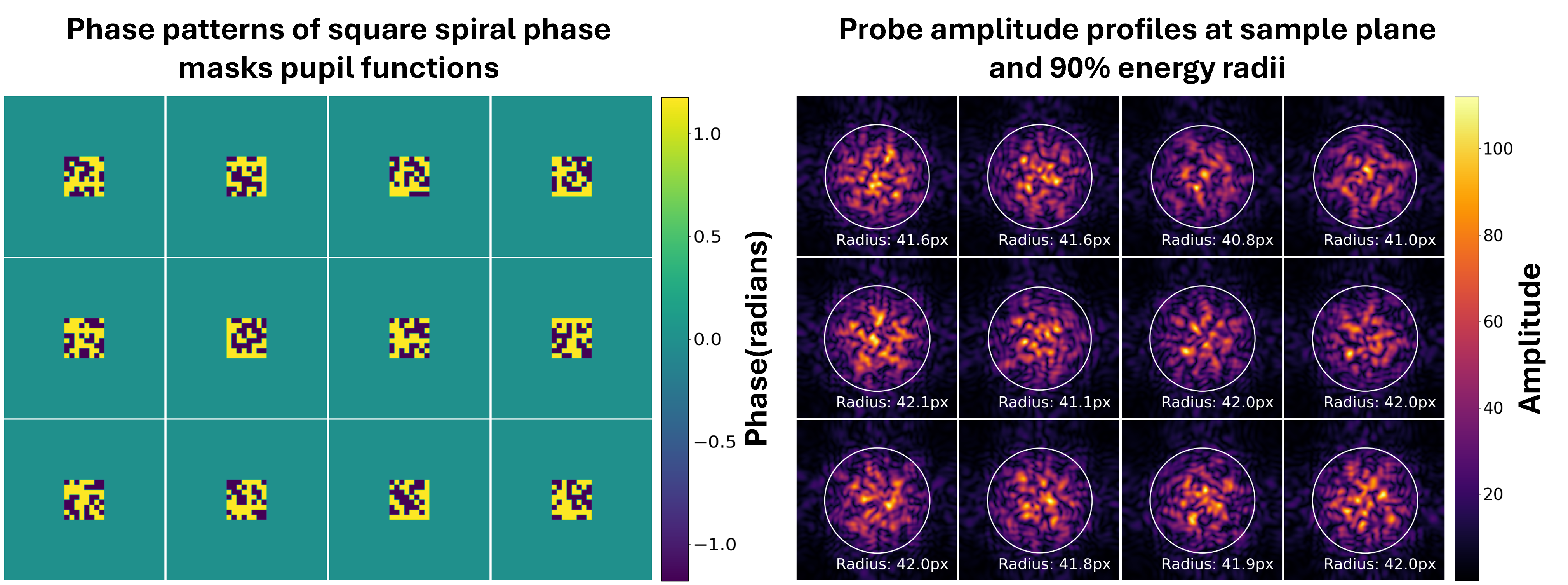}
    \caption{
    Representative examples of probes generated using Hadamard-like binary phase modulations.
    Left: pupil function phase maps. Right: resulting probe amplitudes after propagation to the object plane. 
    Each probe is overlaid with a white circle indicating the effective radius enclosing 90\% of the total intensity. 
    }
    \label{fig:hadamard}
\end{figure}

The amplitude distributions of Hadamard-based probes show excellent spatial uniformity and consistency in size. 
Their effective radii range from 40.8 to 42.1 pixels, with an average of 41.66 pixels and a standard deviation of only 0.42 pixels. 
In addition to their similar sizes, the amplitude distributions are centrally concentrated and exhibit high similarity but yet distinct differences in each probe's speckles, indicating a potential for reliable probe separability and beam uniformity. 
This can be attributed to the binary and highly structured nature of the Hadamard-like phase modulation, which avoids directional bias.
Moreover, this approach is highly feasible, as it can generate a large number of distinct probe patterns by permuting the rows and columns of a Hadamard matrix, without increasing the structural complexity of the underlying phase modulation. 
This means that a large number of distinct probes can be generated with only modest increases in structural complexity, offering strong scalability for high-throughput applications (see Fig.~\ref{fig:hadamard}).

\begin{figure}[htbp]
    \centering
    \includegraphics[width=\textwidth]{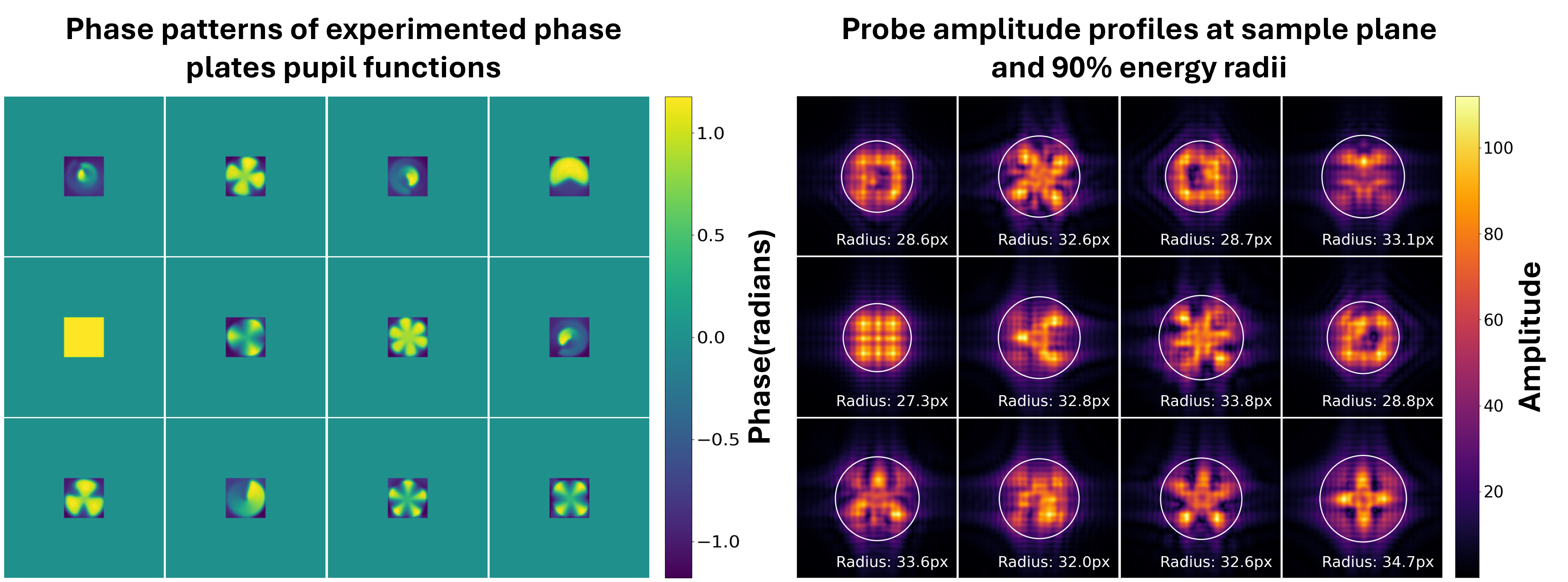}
    \caption{
    Representative examples of probes generated from experimental phase plates.
    Left: pupil function phase maps. Right: resulting probe amplitudes after propagation to the object plane. 
    Each probe is overlaid with a white circle indicating the effective radius enclosing 90\% of the total intensity.
    }
    \label{fig:tang}
\end{figure}

The amplitude distributions of experimentally used phase plate designs display substantial variation in both beam size and spatial uniformity. 
The effective radii span a wide range from 27.3 to 34.7 pixels, yielding an average of 31.56 pixels and a standard deviation of 2.38 pixels. 
Several probes differ significantly from each other in shape and symmetry, producing elongated or irregular amplitude distributions. 
The strong variance of the probes was a deliberate choice to make a set of probes that do not look alike.
Yet, the observed variances in size and shape suggest that, while these designs were practical for fabrication, their performance in terms of uniform multi-beam reconstructability might be less predictable compared to more systematic modulation schemes such as the Hadamard-inspired approach.(see Fig.~\ref{fig:tang}).

\begin{figure}[htbp]
    \centering
    \includegraphics[width=\textwidth]{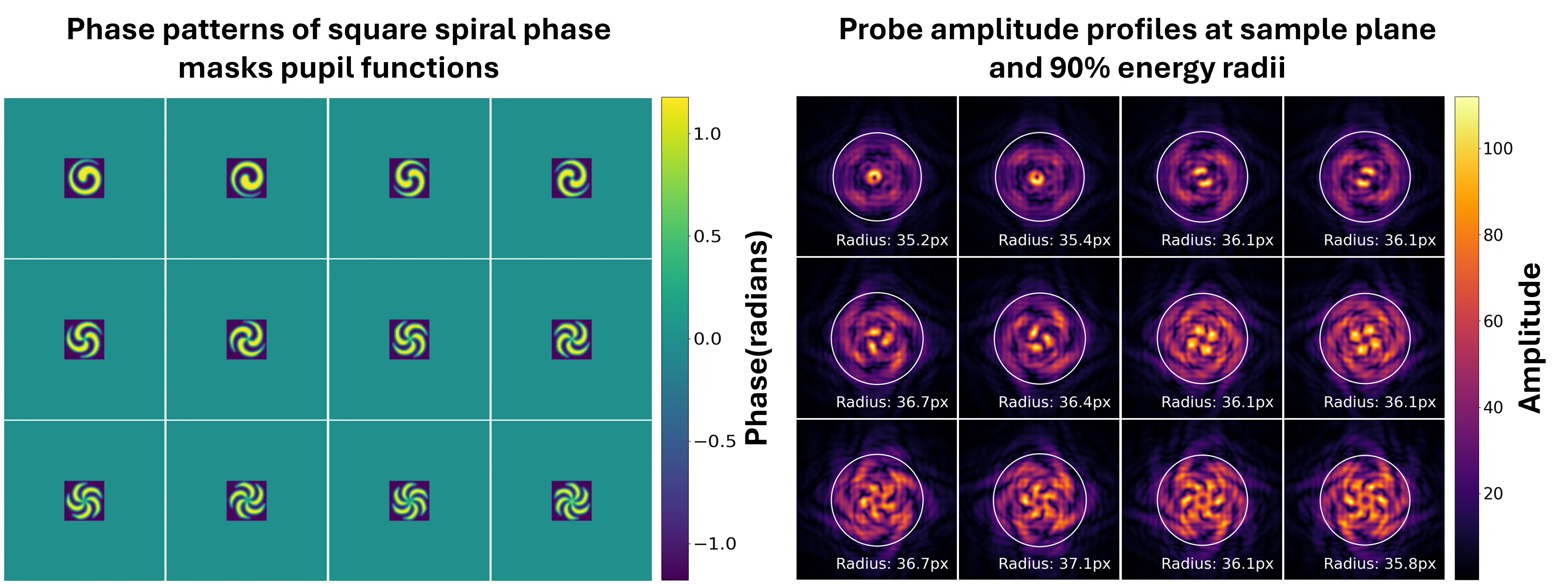}
    \caption{
    Representative examples of using spiral phase masks.
    Left: pupil function phase maps. Right: resulting probe amplitudes after propagation to the object plane. 
    Each probe is overlaid with a white circle indicating the effective radius enclosing 90\% of the total intensity. 
    }
    \label{fig:spiral}
\end{figure}

The pupil functions with spiral phase modulation generate compact, circular beams with relatively stable effective sizes, ranging from 35.2 to 37.1 pixels. 
The average radius is 36.15 pixels, with a standard deviation of 0.5 pixels. 
However, as the number of spiral arms increases, the internal probe structure becomes progressively more complex, with multiple amplitude lobes and higher angular frequencies. 
While rotating the spiral direction allows generating two distinct probes from the same number of arms, scaling to larger probe sets still requires increasing the number of spiral arms.
This leads to a stepwise, approximately linear increase in structural complexity as more probes are added. 
Although spiral phase masks offer consistent beam sizes and can support a modest number of distinct probes, the need for increasingly fine structures limits their scalability and makes fabrication more challenging for large probe arrays.
Although the added complexity does not significantly affect probe size, it imposes practical limitations: generating a large number of distinct probes would eventually require pupil functions with many thin spiral arms, posing manufacturing challenges and reducing scalability for larger probe arrays(see Fig.~\ref{fig:spiral}).

In summary, probes from Hadamard-inspired pupil modulations provide the best uniformity with strong potential for separability among the candidates studied, making them potentially highly suitable for robust multi-beam ptychography. 
Zernike and spiral-based designs offer structured alternatives, though they present trade-offs between potential separability, uniformity, and feasibility constrained by implementation complexity. 
The experimented phase plate designs, while proven to be feasible, require further refinement to meet the probe uniformity critical for consistent overlap ratios across all beams within an MBP dataset.

\subsection{Single-Beam Performance Consistency}

While uniform probe size and shape are important prerequisites for multi-beam imaging, it is equally critical that each individual probe performs equally well when used alone in standard SBP. 
This ensures that all probes contribute meaningfully and consistently to the reconstruction process, especially under varying scan conditions. 

To evaluate how well each probe family performs under standard single-beam ptychographic imaging conditions, we conducted a series of simulations across varying scan position densities. 
For each of the four candidate probe groups, we used all 12 designed probes, as shown in Fig.~\ref{fig:zernike}--\ref{fig:spiral}, to scan the same fixed region of the sample. 
The scan step size was systematically varied, resulting in different numbers of diffraction patterns collected over the same area. 
This approach allowed us to assess both the average imaging performance and the uniformity within each group of probes under consistent object conditions.

\begin{figure}[htbp]
\centering
\includegraphics[width=\textwidth]{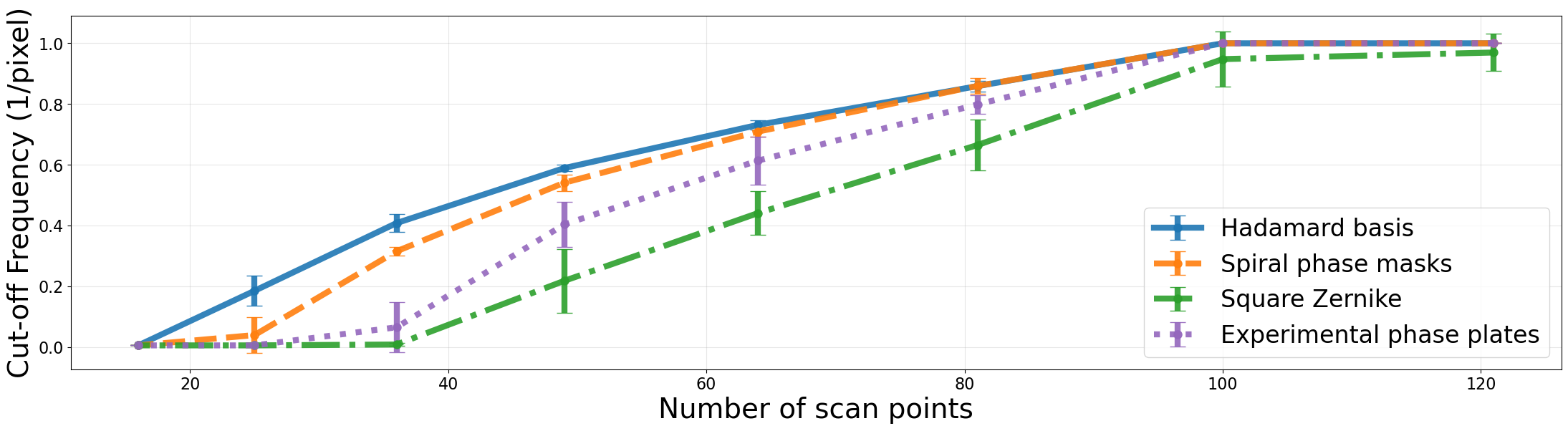}
\caption{
Average single-beam ptychographic reconstruction quality for each probe family as a function of scan position density. 
Each curve represents the mean FRC cutoff frequency, computed using the 1-bit threshold, from 12 individual probes scanning the same sample region at varying step sizes. 
Error bars indicate the standard deviation across the reconstructions from each of the 12 probes.
}
\label{fig:single_performance}
\end{figure}

We again adopted the FRC metric~\cite{van2005fourier} to quantitatively evaluate reconstruction quality, following the same criteria described previously. 
The results are summarized in Fig.~\ref{fig:single_performance}.
We can observe clear differences in how each probe group responds to varying scan position density.
Hadamard-inspired probes not only achieve the highest average resolution across all scan point counts, but also exhibit the lowest standard deviations, highlighting their uniformity and robustness as single-beam illuminations. 
Spiral-phase probes deliver comparable performance to Hadamard at high scan position densities; however, they degrade more rapidly and with larger spreads in quality when the number of scan points is reduced.
Experimented phase plate designs display relatively large variability across scan position densities. 
While their average performance can surpass Zernike designs at intermediate scan densities, they generally remain less consistent than Hadamard or spiral-based probes.
Despite higher scan densities, inconsistent performance across Zernike probes limits their overall reconstruction quality and makes them the least reliable among the tested strategies.

From a practical standpoint, these findings suggest that Hadamard-inspired and spiral-based sets of probes might be more suited for MBP data taking.
In contrast, experimented phase plate designs, while practically feasible and capable of producing reasonable reconstructions, show significant variation in performance across probes and scan conditions, indicating limited reliability for consistent multi-beam imaging.
Zernike-based designs exhibit lower overall reconstruction quality and pronounced variability among individual probes, which becomes more evident at reduced scan densities.

\subsection{Multi-beam Reconstruction Performance}
To evaluate how well the probe designs maintain uniformity and separability when used together, we performed simulations using four-beam configurations.
This assesses the robust combinability of each probe design strategy, meaning whether any randomly selected subset of four probes can function effectively as a multi-beam configuration without performance degradation. 
Given the previously validated uniformity of the object, we adopted a one-to-one mapping between each selected probe within a set of 12 and a corresponding object sub-region to generate the diffraction data for 12 distinct SBP datasets per strategy.  
The latter was repeated for the different previously used scan point densities.
MBP datasets were created by randomly selecting 4 probes from the same family and summing the respective diffraction patterns of the previously simulated SBP datasets.
Repeating this random selection 10 times for each design strategy resulted in 10 different MBP datasets for each probe set and each scan point density.

\begin{figure}[htbp]
\centering
\includegraphics[width=\textwidth]{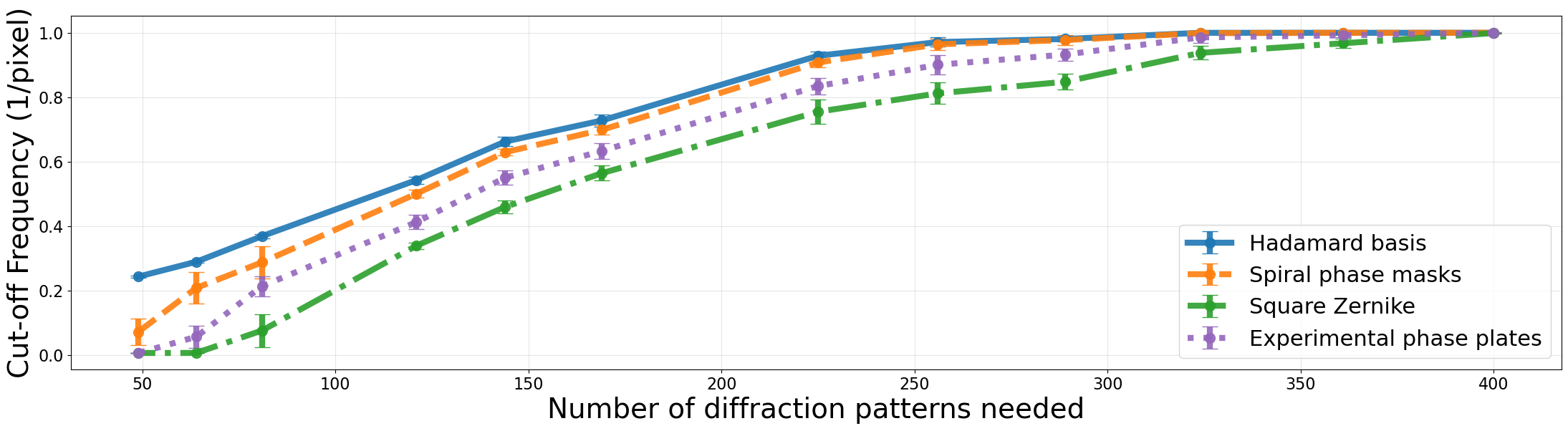}
\caption{
Multi-beam ptychographic reconstruction quality for each probe set as a function of the number of scan points density.
For each group, ten independent four-beam configurations were generated by randomly selecting four probes from the full set of twelve.
Each configuration was used to scan a fixed-size object area at varying step sizes.
FRC-based resolution estimates were calculated independency for each of the four reconstructed regions of every single reconstruction and then averaged.
Error bars indicate the standard deviation across the ten random 4-probe configurations.
}
\label{fig:mbp_results}
\end{figure}

This approach allowed us to quantify reconstruction performance as a function of sampling effort (i.e., number of recorded diffraction patterns in a constant area).  
For each simulation, each of the reconstructed regions corresponding to the four individual beams was evaluated against the ground truth using FRC, and the resulting resolution values were averaged to represent the performance of the selected probe combination in a multi-beam configuration.  
Finally, for each group of ten four-beam configurations, we computed the mean and standard deviation of the resolution across the full range of step sizes.  
These results were plotted as resolution versus the number of diffraction patterns, providing a statistical measure of how well and how consistently each probe design strategy supports multi-beam reconstruction under varying sampling conditions.
Figure~\ref{fig:mbp_results} summarizes the MBP reconstruction performance for each probe family as a function of the number of scan points. 

Among the four candidate probe families, Hadamard-based probes consistently yield the highest resolution across nearly all scan densities, with the smallest standard deviation across different four-beam combinations. 
This strong separability and small variation across random subsets of probes from this strategy are consistent with the previously tested high spatial uniformity of individual probes and high uniformity in reconstructability when used as probes for SBP. 

Spiral phase probes closely follow the Hadamard-inspired group in performance, achieving comparable resolution levels and similarly small variability at higher scan densities.
At lower scan densities, however, their reconstruction quality is slightly lower than that of Hadamard, and a more noticeable variation emerges across different probe combinations.
Despite the need for increasingly fine structures at higher orders, spiral-based probes maintain consistent beam sizes and deliver robust performance in MBP configurations, making them a practical and effective alternative.

Experimented phase plate designs show lower resolution than Hadamard and spiral-based probes, though their average performance always surpasses Zernike-based probes.
While individual reconstruction quality varies due to irregular spatial features and size differences, the variation across different probe combinations remains relatively modest.
This consistency is likely a result of the deliberate design of each phase plate to ensure mutual distinguishability.
However, the lack of geometric regularity still limits their overall scalability and reliability in MBP applications, particularly compared to more structured modulation strategies.

Square Zernike probes exhibit the lowest overall resolution among the tested families, and show notable variation across configurations. 
While the average resolution improves steadily with increased sampling, the variability reflects a lack of uniformity and separability in certain combinations. 
This suggests that although mathematically orthogonal, square Zernike probes may require further selection of good subsets for optimal MBP use.

Taken together, these results highlight the importance of structured design strategies for MBP, particularly those that ensure spatial uniformity, uniform SBP performance, and still allow for the separability across probe subsets.

\section{Conclusion}

Ptychographic imaging is highly sensitive to the choice of the illuminating probe.
This remains essential in MBP, where multiple beams must not only perform well individually but also function effectively as a set of multiple probes. 
Our study shows that using multiple individually well-performing probes is insufficient to guarantee high-quality MBP reconstructions.
The performance of the entire probe set together is crucial for accurate and efficient MBP reconstructions.

We systematically evaluated four representative probe design strategies: Zernike polynomials adapted to square aperture, Hadamard-like binary phase patterns, previously experimented phase plates, and spiral phase masks. 
Using consistent performance metrics and simulation protocols, we found that Hadamard-based probes fulfill all key design requirements: they exhibit strong separability, high uniformity across scan conditions, and feasibility.
They have excellent scalability and high photon efficiency. 
In addition, their 2D binary phase-only structure makes them experimentally feasible and easy to fabricate, enabling the creation of large and well-behaved probe sets.

Spiral-based designs yield similarly high reconstruction quality and low variability, but their scalability is limited since the required feature size decreases with the number of beams. 
In contrast, Zernike-based and the set of experimental phase plates commonly used in MBP resulted in worse reconstruction qualities, especially towards lower scan point densities, with a higher variance for equal scan parameters.

These findings highlight the importance of designing probes that are not only high-performing individually but also collectively optimized for separability, uniformity, and feasibility. 
As MBP systems scale to larger beam counts and more complex imaging scenarios, future work should focus on systematic strategies to optimize both separability and uniformity across high-dimensional probe sets. 
This includes exploring new orthogonal basis functions, robust fabrication methods for structured phase elements, and experimental protocols that validate simulated performance under realistic conditions. 
Such efforts will be critical in translating design principles into scalable and reliable MBP implementations.
Out of the tested strategies, both the Hadamard-like and spiral-based ways of structuring pupil functions seem the most suited approaches to create large sets of probes for MBP.

\section*{Acknowledgements}

This work was supported by the Röntgen-Ångström Cluster (RÅC) grant (VR 2021-05975). 
Computations were performed using resources provided by the MAX IV Laboratory. 
We acknowledge MAX IV Laboratory, a Swedish national user facility, supported by Vetenskapsrådet (Swedish Research Council, VR) under contract 2018-07152, Vinnova (Swedish Governmental Agency for Innovation Systems) under contract 2018-04969, and Formas under contract 2019-02496.

\section*{Disclosures}

The authors declare no conflicts of interest. 
The simulated raw data and the reconstruction scripts, including configuration files required to reproduce the Ptypy simulations, will be made available in a public repository upon publication.

\bibliography{sample}

\end{document}